\documentclass{cimento}

\usepackage{graphicx,url,paralist}

\title{Gamma-ray, Particle and Exotic Physics at TeV energies with the MAGIC telescopes}
\author{Michele Doro\\University and INFN Padova, via Marzolo 8,
  I-35131 Padova, Italy}

\begin{document}
\maketitle

\begin{abstract}
  MAGIC is an instrument composed of a pair of telescopes for
  gamma-ray and cosmic-ray astrophysics in the TeV range. It is
  operating for more than a decade now, and is one of the current best
  performing instruments in this field, specifically at low energies,
  where it achieves the largest sensitivity. MAGIC pursues a strong program in
  galactic and extragalactic gamma-ray science. Its catalog of blazars,
  radiogalaxies and galaxy clusters observations as well as supernovae,
  novae, pulsar wind nebulae, pulsars and binary systems has now
  increased to several tens of detected targets. In addition, MAGIC is suited
  for cosmic ray searches, being sensitive to the signatures of
  earth-skimming tau-neutrinos, cosmic antiprotons, and others. Furthermore,
  MAGIC has a strong fundamental physics program, with searches for
  particle dark matter, Lorentz Invariance violations, axion-like
  particles and primordial black hole evaporation, providing important
  recent constraints in some relevant cases. Finally, MAGIC has a
  follow-up program of Gravitational Waves events. Few highlights
  topics will be discussed in this contribution.s
\end{abstract}

\paragraph{\bfseries{Introduction}}
MAGIC (Major Atmospheric Gamma-ray Imaging Cherenkov) is a telescope
composed of a couple of 17-m diameter reflector of the IACTA (Imaging Atmospheric
Cherenkov Telescope Array) class. The technique is based on the
observation of the Cherenkov
radiation generated in extended particle showers produced in the
atmosphere by either cosmic particles of high energy cosmic
radiation reaching the Earth. The sampling of the Cherenkov radiation guarantees good
sensitivity, low energy threshold, good energy and angular resolutions,
compared to shower-front particle detectors, at the expense of a duty cycle
limited to dark or moderate moonlit nights. MAGIC is located at
the Canary Island of La Palma (Spain), in the northern hemisphere, at
the Observatorio de Roque de Los Muchachos, which will be soon hosting
the new IACTA generation of the CTA project. The energy working range
is 50 GeV to 50 TeV, with a sensitivity of 0.66\% of the Crab Nebula
flux above 220~GeV in 50~h~\cite{Aleksic:2014lkm} and the instruments
is in operation for a decade now.

The MAGIC science program is wide (see sketch figure), and focused on VHE
gamma-ray astrophysics. This encompasses relativistic jets physics,
plasma physics, magnetic reconnection, diffuse shock accelerations,
cosmic ray transport 
and so on, requiring often strong interdisciplinarity.
The science pursues can be done in multiple galactic and
extragalactic targets and target classes. This will be discussed in
Sec.~1. However, with non standard reconstruction 
algorithm, MAGIC can be operated as a particle detector, in the sense
that it can discriminate cosmic particles: protons, electrons,
etc.
This is the topic of Sec.~2.
Finally, MAGIC has
the characteristics to be sensitive to some New Physics scenarios,
including dark matter and Lorentz Invariance violation, but also (if
existing) to electromagnetic counterparts of gravitational wave
events.
These
will be discussed in Sec.~3 and 4
respectively.

Due to space limitation, this contribution will be a brief overview. More detailed reports on MAGIC physics and
the success of IACTA can be found for example in~\cite{Lorenz:2012nw,Aharonian:2008zza}.  

\begin{figure}[t]
  \centering
  \includegraphics[height=7cm]{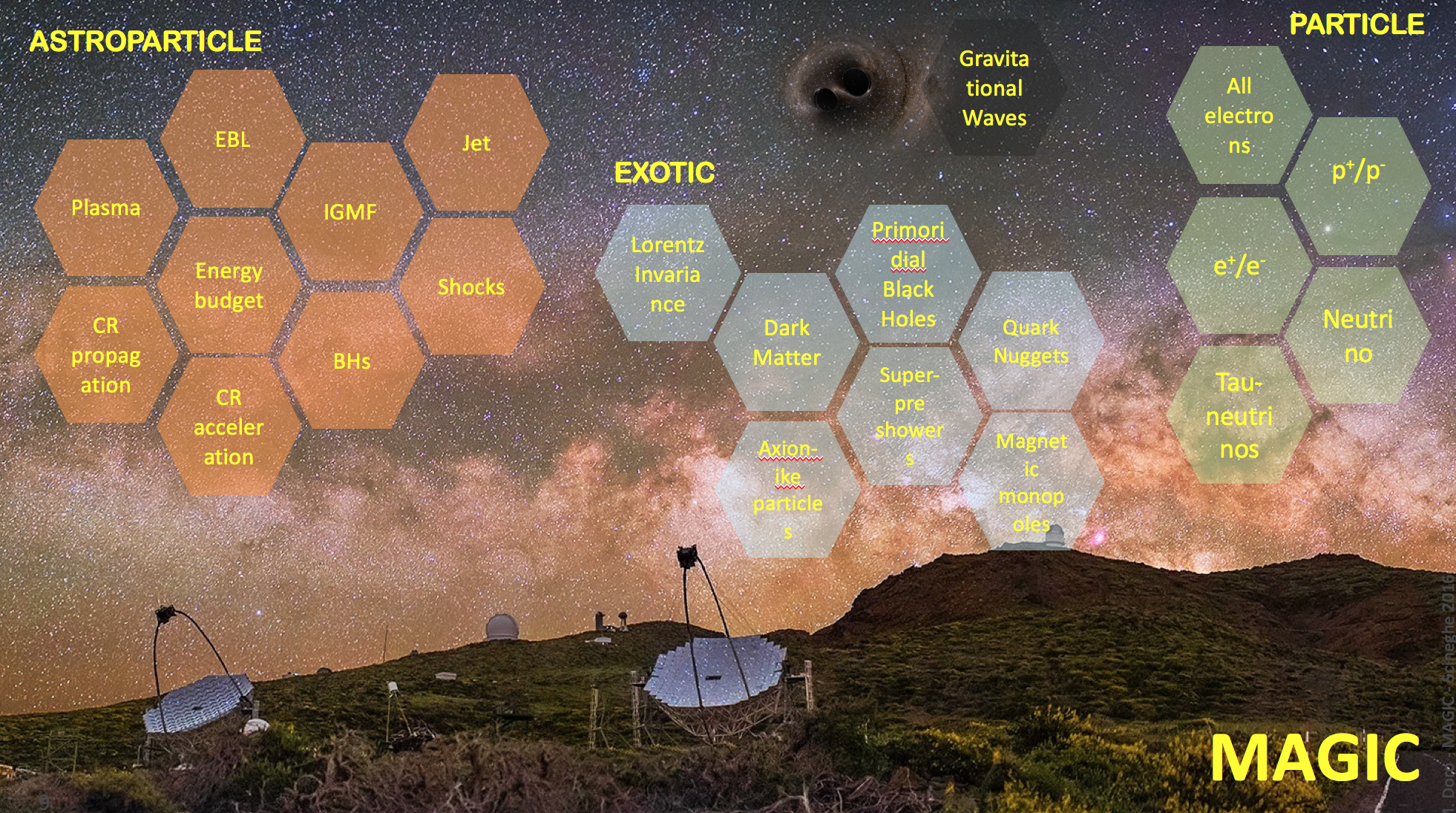}
  \caption{Science cases with very high energy gamma rays with MAGIC\label{fig:gamma}}
\end{figure}

\paragraph{\bfseries{1. MAGIC as a TeV gamma-ray detector}}
\label{sec:gamma}
The presence of VHE (Very High Energy: $>100$~GeV) gamma rays in
astrophysical environment is always connected to 
parent highly accelerated cosmic ray particle (leptons and/or
hadrons). Therefore gamma ray astrophysics probes cosmic ray
physics. However, to be able to understand the particle and radiation fields in
astrophysical environments, a large astronomical framework of knowledge is
required, formed by a wide multi-wavelength and multi-instrument contribution. Classical targets for
VHE gamma-ray observations were mentioned in the abstract.
It is impossible to summarize the many results achieved with 
observation at VHE in the past decade. Only a selection will be
discussed, in which MAGIC has shown unique capabilities. 

The low MAGIC energy threshold (50 GeV), had allowed to search for the
farthest sources in the sky, because of the opaqueness of the Universe
to VHE gamma rays lost due to pair production with the UV-IR background
radiation fields of the Extragalactic Background Light (EBL), the
second radiation field permeating the universe after the Cosmic
Microwave Background (CMB). MAGIC has recently expanded the TeV
universe by doubling the distance of the farthest sources known. Two
sources at redshift $z\sim 1$ were detected~\cite{Ahnen:2015qrv,Ahnen:2016vlr}: the galaxy PKS 1441+25 at
$z=0.939$ and the galaxy B0218+35 at $z=0.944$
of which MAGIC did not observe the direct light, but that
gravitationally lensed from an intervening spiral galaxy in
between. This was the first gravitational-lensed delayed VHE signal
ever observed.

The importance of multi-wavelengths campaign has become utter,
and indeed MAGIC has developed several monitoring campaigns and Time
of Opportunity programs based on external triggers. For example, during the Mrk421 campaign
reported in \cite{Balokovic:2015dnz}, the source had the coverage of
Swift-UVOT, Swift-XRT, NuSTAR, Fermi-LAT, MAGIC, and VERITAS. The
flare event not only generated an increased flux, but also shifted the
emission peak toward higher energies. In the Synchroton Self-Compton
scenarios, this may hint to a different (than baseline) electron
population that was swept-up or one single electron population
that was boosted for some mechanism. The energy-dependent variability provided
an important part of the puzzle of the global picture of the target
dynamics.
%
The flares in Active Galactic Nuclei (AGNs) provide useful insights
into the associated Black Hole (BH) 
physics. All kinds of AGNs in MAGIC catalog have shown extremely fast variability.
In 2014, MAGIC saw an impressive flare of the radio-galaxy IC 310. The
flux doubled in 4.8 minutes, from an object located at 78~Mpc. It is
very unclear what mechanisms could provide such boost because
rise-time constraints emission size: the emission region must have a size
smaller than the 20\% of the BH size. The proposed explanation is that particle
acceleration occurs by the electric field across a magnetospheric gap at the
base of the radio jet, much alike the pulsar radiation
mechanism~\cite{Aleksic:2014xsg}.  

Galaxy clusters are expected to show a diffuse gamma-ray emission due
to the interaction of accelerated CR with the ambient intracluster
medium. Perseus (78~Mpc far) is a cool-core clusters, the
brightest in X-ray and thus an optimal lab to search for gamma rays
associated to accelerated particle fields in the core. The explanation for the origin of radio halos is in fact
challenging and TeV gamma-ray are an unique probe. TeV gamma 
rays are expected in hadronic model because radio-emitting electrons are
secondaries produced by CR protons interacting with the protons of the
ICM, or in re-acceleration model where seed population of CR electrons
are re-accelerated by interacting turbulent waves. MAGIC observed
for 250 h (selected) in 4 years, providing several (model-dependent)
constraints. With MAGIC data, we
constrained the fraction of the energy dissipated in structure
formation shocks that goes into particle acceleration to be not larger
than
37\%. We limited the cosmic-ray to thermal-pressure ratios to values
between 2 and 20\%. We constrained the magnetic fields that produce
the observed synchrotron emission from secondary electrons to be
smaller than 10 $\mu$G~\cite{Ahnen:2016qkt}.

Other recent MAGIC results are available here: \url{https://magic.mpp.mpg.de/backend/results/latest}.
  
\paragraph{\bfseries{2. MAGIC as a TeV particle detector}}
\label{sec:particle}
Besides cosmic gamma rays, atmospheric showers are initiated by cosmic
rays. Actually, the number of showers generated by the latter class is
more than a factor of thousand larger than those initiated by gamma
rays. The ability of IACTA to discriminate between gamma and
particles, and among particles, stems mostly from the different
imprint in the camera produced by different primaries.
MAGIC is routinely used as a particle detector. It can perform
all-electron searches~\cite{BorlaTridon:2011dk,Mallot:tevpa}, but in
principle can also discriminate the charge of the particle by using
the shadow of the Moon~\cite{Colin:2011wc}. MAGIC also follows-up
triggers from High Energy events from the IceCube neutrino detector~\cite{Aartsen:2016qbu}.

In the following, we focus on the preliminary results MAGIC
achieved on the capability to discriminate cosmic tau-neutrinos events
in the data, thus opening the possibility of MAGIC as a neutrino
detector. This challenging method 
is based on the fact that MAGIC, from its mountain location, has a
``window of visibility'' of $80\times5$ deg$^2$ in the direction of the
Atlantic Ocean, at zenith angles slightly below $90^\circ$. The Ocean
is about 170~km away. By pointing the
telescope in this window, MAGIC could detect atmospheric shower
generated by tau-leptons emerging from the ocean's surface, and
soon-after decaying and generating particle showers. In turn, this very energetic tau-lepton (the
accessible energy range is beyond the PeV), must
have been produced not too deep inside the ocean by the conversion of
a very energetic cosmic tau-neutrino entering the Earth in the same
MAGIC observing direction. The technique is discussed
in~\cite{Gora:2016mmy} (and references therein). Tau-induced showers
are discriminated because they are created much closer to the
telescope compared to any other possible source of
background~\cite{MAGIC:2017tau}. Although the sensitivity to these 
signals is in principle rather high for MAGIC, at the level of even
above that of IceCube above the PeV range, the prospects for actual
detection of cosmic tau-neutrinos are still thin, due to very low
expected flux. This maybe not be the case in the occurrence of
a strong flare (of an AGN, or a gamma ray bursts, or from a
gravitational wave event) where hadrons are expected to be strongly accelerated, thus
providing also an ample neutrino flux. In
this case, if the source passed through the ``ocean window'', the
neutrino flux can be indirectly observed. This observation mode is
still under development in MAGIC. However, even 
null observation from a flaring source could provide interesting upper
limits on the amount of tau-neutrino produced locally,
competitive to other instruments in the field.

\paragraph{\bfseries{3. MAGIC as a New Physics detector}}
There are many theoretical arguments indicating that there may be
new Physics around the TeV scale. IACTA could be valuable probes for
Lorenz Invariance Violations, by observing the time-of-flight delays
of photons at difference energy, a tiny effect possibly amplified by
the cosmological distances at hand~\cite{Albert:2007qk}; IACTA can probe Axion-like
Particles (ALPs), that may explain the otherwise awkward observation of TeV photons
from targets where we expect photons should be hardly observable, because
of absorbing medium either located at the source (e.g. dense gas
fields) or in the path to the source. By converting to ALP, faraway
TeV photons could thus travel larger distances than expected.
IACTA can see gamma-ray flashes from fast
evaporating Black Holes generated in the early Universe and
evaporating in these times; finally, the passage of exotic objects
like magnetic monopoles or quark nuggets, could also generate
Cherenkov light observable with MAGIC (see recent review~\cite{Doro:rare}).

However, large attention is paid to IACTA for their capability as dark
matter (DM) detectors. They are indeed valuable DM probes, for the
following reasons: $a)$ it is 
often the case that gamma rays are found in prompt annihilation or
decay reaction in DM rich environment, $b)$ the associated
gamma ray spectrum, for many DM scenarios, can peak at the TeV and may present peculiar features
that can easily differentiate it from a purely astrophysical one, $c)$
DM spectra would be universal, so several different targets
may be observed with exactly the same spectra, $d)$ the shape of the
spectra and its cutoff at the DM mass provide also clues for
identification of DM besides detection. DM searches
are done with MAGIC at several targets (see a recent
review~\cite{Doro:2017dm}). The cleanest targets in the sky are the
dwarf galaxies gravitationally bound to the Milky Way and orbiting 
its the DM halo. These are small galaxies, with almost no stellar
activity, where DM must have accreted in large concentrations
(mass-to-light ratios thousands of times that of the Sun), as seen
by stellar velocity dispersion. MAGIC has a large track in observing
these objects. The recent large campaign (160~h) on the Segue~1 satellite
galaxies provided the strongest constraints from this class of targets
above few hundreds of GeV, and received the attention of the Particle
Data Group~\cite{DPG}.

\label{sec:exotic}
\paragraph{\bfseries{4. MAGIC as a Gravitational Waves counterpart detector}}
\label{sec:gw}
Gravitational waves (GWs) associated physics has recently gained a strong
interest due to multiple detections of GWs by the Ligo Virgo
Collaboration (LVC). Although the theoretical mapping is far
from being complete, it is believed that if the merger event happen
between at least one compact object (NS-NS or NS-BH mergers), a measurable
gamma-ray counterpart may be expected, although its spectral energy
distribution is unclear. MAGIC can rapidly reposition ($<20$~sec)
after receiving a GW alert from LVC. However, its small field of view
(about 10 square degrees) cannot sample but a small fraction of the
LVC region of interest. Despite so, MAGIC has performed follow-ups of
GW161226~\cite{gw} and is refining its strategies toward a better
selection of pointings.

\paragraph{{\bfseries Conclusion}}
Although not the mainstream science, among the MAGIC science goals is
that of trying to add a piece in the puzzle of cosmic ray fluxes at
Earth, hunting elusive tau-neutrinos, finding signatures of New
Physics. The astrophysical results are disparate, and from tens of
of targets and target classes. Result are now contributing to
astrophysics in a wide multi-instruments and multi-wavelengths
scenarios, placing gamma-ray astrophysics into the realm of a true
astronomical field. MAGIC has contributed in many ways to this success
along the past 12 years. Few years ahead will be productive until the
times come for the future generation of instruments in this field,
with the Cherenkov Telescope Array observatory~\cite{CTA:2010bc} already in construction
close to MAGIC at the island of La Palma.



\begin{thebibliography}{0}

\bibitem{Aleksic:2014lkm}
  MAGIC Coll.,
  Astropart.\ Phys.\  {\bf 72} (2016) 76
\bibitem{Lorenz:2012nw}
  E.~Lorenz and R.~Wagner,
  Eur.\ Phys.\ J.\ H {\bf 37} (2012) 459
\bibitem{Aharonian:2008zza}
  F.~Aharonian {\it et al.},
  Rept.\ Prog.\ Phys.\  {\bf 71} (2008) 096901.
\bibitem{Ahnen:2015qrv}
  MAGIC and Fermi-LAT Colls.,
  ApJ  {\bf 815} (2015) no.2,  L23
\bibitem{Ahnen:2016vlr}
  MAGIC Coll.,
  A\&A {\bf 595} (2016) A98
\bibitem{Balokovic:2015dnz}
  VERITAS, MAGIC Colls. and NuSTAR,
  ApJ  {\bf 819} (2016) 156
\bibitem{Aleksic:2014xsg}
  MAGIC Coll.,
  Science {\bf 346} (2014) 1080
\bibitem{Ahnen:2016qkt}
    MAGIC Coll.,
  Astron.\ Astrophys.\  {\bf 589} (2016) A33
\bibitem{BorlaTridon:2011dk}
  D.~Borla Tridon {\it et al.},
  Procs. ICRC (2011)
\bibitem{Mallot:tevpa}
  K. Mallot {\it et al.},
  TeVPa Conference, Tokyo, (2015)
  \bibitem{Colin:2011wc}
  P.~Colin {\it et al.},
  Procs. ICRC (2011)
\bibitem{Aartsen:2016qbu}
  IceCube, MAGIC and VERITAS Colls.,
  JINST {\bf 11} (2016) P11009
\bibitem{Gora:2016mmy}
  D.~Gora and E.~Bernardini,
  Astropart.\ Phys.\  {\bf 82} (2016) 77
\bibitem{MAGIC:2017tau}
  D. Gora, AIP Conference Proceedings {\bf 1792} (2017) 060008
  In prep.
\bibitem{Albert:2007qk}
  MAGIC Coll., {\it et al},
  Phys.\ Lett.\ B {\bf 668} (2008) 253
\bibitem{Doro:rare}
  M.~Doro,
  EPJ Web Conf. {\bf 136} (2017) 01003 
\bibitem{Doro:2017dm}
  M.~Doro,
  Procs. of the 25th European Cosmic Ray Symposium (ECRS 2016), 04-09
  September 2016. Turin, Italy. CNUM: C16-09-04.3
\bibitem{DPG}
  C. Patrignani {\it et al.} (Particle Data Group),
  Chin. Phys. C, {\bf 40} (2016) no.10, 100001
\bibitem{gw} GCN \#18776
\bibitem{CTA:2010bc}
  CTA Consortium,
  Exper.\ Astron.\  {\bf 32} (2011) 193
\end{thebibliography}
\end{document}